 \documentclass[final,3p,times,twocolumn]{elsarticle}

 \usepackage{graphicx}

\usepackage{amssymb}

\journal{Physica B}
\begin{document}

\def\tc{$T_{c}$\,}

\def\t6as{$\mathrm{(TMTSF)_{2}AsF_{6}}$}
\def\tmx{$\mathrm{(TMTSF)_{2}(ClO_{4})_{(1-x)}ReO_{4}_x}$\,}
\def\tmc{$\mathrm{(TMTSF)_{2}ClO_{4}}$\,}
\def\tms{$\mathrm{(TMTSF)_{2}AsF_{6(1-x)}SbF_{6x}}$}\,
\def\tmps{$\mathrm{(TMTTF)_{2}PF_{6}}$\,}
\def\tmttfsbf6{$\mathrm{(TMTTF)_{2}SbF_{6}}$\,}
\def\tmttfasf6{$\mathrm{(TMTTF)_{2}AsF_{6}}$\,}
\def\tmttfbf4{$\mathrm{(TMTTF)_{2}BF_{4}}$\,}
\def\tmtsfreo4{$\mathrm{(TMTSF)_{2}ReO_{4}}$\,}
\def\tq{$\mathrm{TTF-TCNQ}$\,}
\def\tsq{$\mathrm{TSeF-TCNQ}$}\,
\def\qnq{$(Qn)TCNQ_{2}$}\,
\def\R{$\mathrm{ReO_{4}^{-}}$}  
\def\C{$\mathrm{ClO_{4}^{-}}$}
\def\P{$\mathrm{PF_{6}^{-}}$}
\def\tqr{$\mathrm{TCNQ^\frac{\cdot}{}}$\,}
\def\nmpq{$\mathrm{NMP^{+}(TCNQ)^\frac{\cdot}{}}$\,}
\def\q{$\mathrm{TCNQ}$\,}
\def\nmp{$\mathrm{NMP^{+}}$\,}
\def\f{$\mathrm{TTF}\,$}
\def\tc{$T_{c}$\,}
\def\nmq{$\mathrm{(NMP-TCNQ)}$\,}
\def\ts{$\mathrm{TSeF}$}
\def\tsm{$\mathrm{TMTSF}$\,}
\def\tst{$\mathrm{TMTTF}$\,}
\def\tmp6{$\mathrm{(TMTSF)_{2}PF_{6}}$\,}
\def\tms2x{$\mathrm{(TMTSF)_{2}X}$}
\def\as{$\mathrm{AsF_{6}}$}
\def\sb{$\mathrm{SbF_{6}}$}
\def\pf{$\mathrm{PF_{6}}$}
\def\re{$\mathrm{ReO_{4}}$}
\def\ta{$\mathrm{TaF_{6}}$}
\def\cl{$\mathrm{ClO_{4}}$}
\def\4fb{$\mathrm{BF_{4}}$}
\def\ttdm{$\mathrm{(TTDM-TTF)_{2}Au(mnt)_{2}}$}
\def\edt{$\mathrm{(EDT-TTF-CONMe_{2})_{2}AsF_{6}}$}
\def\tfx{$\mathrm{(TMTTF)_{2}X}$\,}
\def\tsx{$\mathrm{(TMTSF)_{2}X}$\,}
\def\tmx{$(TMTSF)_{2}(ClO_{4})_{(1-x)}(ReO_{4})_{x}$\,}
\def\ttftcnq{$\mathrm{TTF-TCNQ}$\,}
\def\ttf{$\mathrm{TTF}$\,}
\def\tcnq{$\mathrm{TCNQ}$\,}
\def\bedtttf{$\mathrm{BEDT-TTF}$\,}
\def\reo4{$\mathrm{ReO_{4}}$}
\def\bedtttfreo4{$\mathrm{(BEDT-TTF)_{2}ReO_{4}}$\,}
\def\et2i3{$\mathrm{(ET)_{2}I_{3}}$\,}
\def\et2x{$\mathrm{(ET)_{2}X}$\,}
\def\ket2x{$\mathrm{\kappa-(ET)_{2}X}$\,}
\def\cuncnbr{$\mathrm{Cu(N(CN)_{2})Br}$\,}
\def\ket2x{$\mathrm{\kappa-(ET)_{2}X}$\,}
\def\cuncncl{$\mathrm{Cu(N(CN)_{2})Cl}$\,}
\def\cuncs{$\mathrm{Cu(NCS)_{2}}$\,}
\def\betsfecl4{$\mathrm{(BETS)_{2}FeCl_{4}}$\,}
\def\bets{$\mathrm{BETS}$\,}
\def\hc2{$H_{c2}(T)$}
\def\et{$\mathrm{ET}$\,}
\def\tmm{$\mathrm{TM}$\,}
\def\tmtsf{$\mathrm{(TMTSF)}$\,}
\def\tmttf{$\mathrm{(TMTTF)}$\,}
\def\tm2x{$\mathrm{(TM)_{2}X}$\,}
\def\t1{${1/T_1}$\,}

\begin{frontmatter}




\title{Towards a consistent picture for quasi-1D organic superconductors}


\author[lab1]{N. Doiron-Leyraud}
\author[lab2]{P. Auban-Senzier}
\author[lab1]{S. Ren\'e de Cotret}
\author[lab3]{K. Bechgaard}
\author[lab2,CIFAR]{D. J\'erome\corref{}}
\ead{jerome@lps.u-psud.fr}
\cortext[cor5]{}
\author[lab1,CIFAR]{L.Taillefer}

\address[lab1]{D\'epartement de physique and RQMP, Universit\'e de Sherbrooke, Sherbrooke, Qu\'ebec, J1K 2R1 Canada}

\address[lab2]{Laboratoire de Physique des Solides, UMR 8502, CNRS - Univ. Paris-Sud, B\^at. 510,  91405,Orsay, France}

\address[lab3]{Department of Chemistry, H.C. {\O}rsted Institute, Copenhagen, Denmark}

\address[CIFAR]{Canadian Institute for Advanced Research, Toronto, Ontario, M5G 1Z8 Canada}

\begin{abstract}
The electrical resistivity of the quasi-1D organic superconductor \tmp6 was recently measured at low temperature from the critical pressure needed to suppress the 
spin-density-wave state up to a pressure where superconductivity has almost disappeared\cite{Doiron09}. This data revealed a direct correlation between the onset of superconductivity at $T_c$ and the strength of a non-Fermi-liquid linear term in the normal-state resistivity, going as $\rho(T) = \rho_0 + AT + BT^2$ at low temperature, so that $A \to 0$ as $T_c \to 0$. Here we show that the contribution of low-frequency antiferromagnetic fluctuations to the spin-lattice relaxation rate is also correlated with this non-Fermi-liquid term $AT$ in the resistivity. These correlations suggest that anomalous scattering and pairing have a common origin, both rooted in the low-frequency antiferromagnetic fluctuations measured by NMR. A similar situation may also prevail in the recently-discovered iron-pnictide superconductors. 

\end{abstract}

\end{frontmatter}

 Organic superconductivity has been discovered in the organic salt  \tmp6  which belongs to the much broader isostructural family of \tm2x compounds    where \tmm
stands for the molecules \tmtsf or \tmttf. This diagram has been studied extensively over the last 30 years with detailed reports in the literature\cite{Jerome91}.
However, the electronic transport of the metallic state has attracted much attention since the finding of a strongly temperature dependent resistance prior to the SC transition around 1K was  unusual for a metal as the resistance is expected to saturate at low temperature following a power law with an exponent larger than unity.
In the initial discovery a sublinear temperature profile of the resistivity in the vicinity of the superconducting transition  had been noticed\cite{Jerome80} and attributed to precursors of the  SC transition itself\cite{Schulz81}. Above this precursor regime, the resistivity remained quasi-linear up to about 10K, a temperature above which a more traditional  quadratic Fermi liquid-like regime was recovered. This  behaviour proportional to $T$  in the low temperature limit has also been  recognized by  other groups \cite{Murata82} and considered as an important experimental feature in these materials.
The fact that a strong temperature dependence of the resistivity prevails at low temperature had thus been considered  as a  piece of evidence for the existence of  paraconductivity in a quasi-one dimensional conductor above $T_c$  \cite{Schulz81}. The calculation of  Azlamazov-Larkin diagrams based on a time dependent Ginzburg-Landau theory provided, (i) a 3D domain restricted to the very vicinity of $T_c$ where the transverse coherence extends over several interchain distances $d$, \textit{ i.e}, $\xi_{\perp}/d>1$ with $\rho_{\parallel}^{3D}$ $ \alpha$ $ T^{-1/2} (ln T/T_c)^{1/2}$ leading in turn to the classical $1/(T-T_c)^{1/2}$ divergence of the paraconductivity and (ii) a 1D regime with $\xi_{\perp}/d<1$ with $\rho_{\parallel}^{1D}$ $ \alpha$ $ T^{1/2} (ln T/T_o)^{-3/2}$ where $T_o$ represents a renormalized mean field temperature, about one third  of the actual mean field temperature \cite{Jerome82}. 
\begin{figure}[ht]
\centerline{\includegraphics[width=1\hsize]{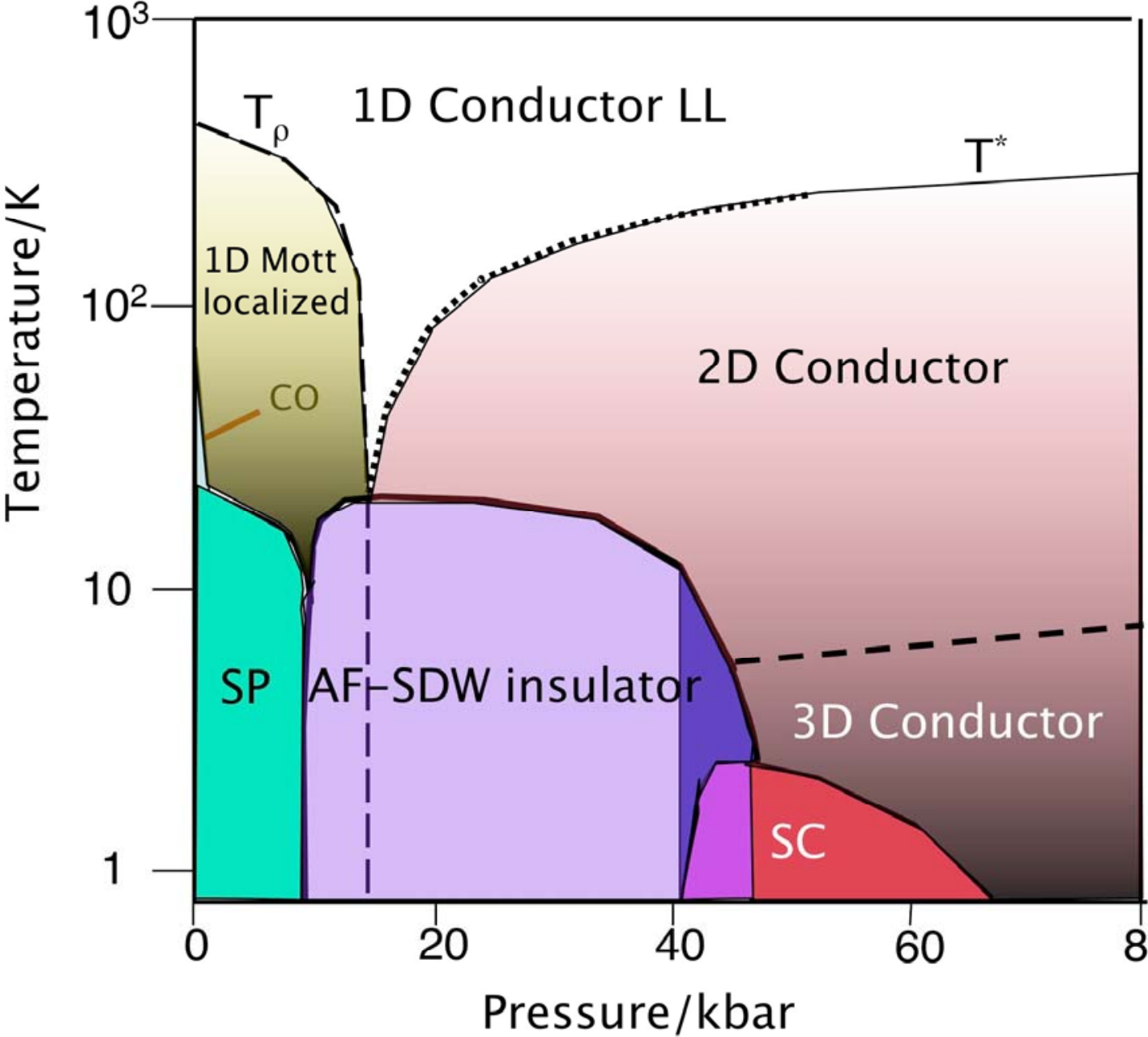}}
\caption{Generic temperature-pressure phase diagram of the  \tm2x family of organic conductors. The diagram is drawn for the compound \tmps taken for the origin of the pressure scale. The location of the compounds \tmp6 and \tmc  under ambient pressure are $\approx37$ and 47 kbar respectively in this diagram.} 
\label{Figure1.pdf}
\end{figure}

While the comparison between  experiments and the theory for paraconducting fluctuations in  ref\cite{Schulz81} revealed a reasonable agreement in the 3D regime. 
 The fluctuation model  failed to reproduce the correct temperature dependence in the range  4-15K  and revisiting this problem appeared to be required.

Furthermore,  magnetism is also  an important feature in the generic phase diagram of \tm2x compounds since a good nesting of the quasi-1D  Fermi surface \tmp6 leads to a spin density wave transition at 12K accompanied by the opening of a gap over most of the Fermi surface. Increasing  pressure   degrades nesting and  gives way to SC, first inhomogenous at 6 kbar\cite{Kang09} and subsequently homogenous above $P_c$ $\approx$ 9 kbar\cite{Jerome04}, (figure\ref{Figure1.pdf}).  The SDW instability can be revived under magnetic field applied along the direction of weakest interchain coupling and fairly well understood in terms of weak coupling theories\cite{Heritier84,Gorkov84}.
An additional evidence for the role of repulsive interaction leading to magnetic ordered ground states came from the strong enhancement of the NMR relaxation rate, \t1\cite{Creuzet85,Bourbonnais88,Brown08} revealing the existence of strong antiferromagnetic spin fluctuations  in the metallic state above the superconducting instability. These features   raise relevant questions about the interplay between pairing and the existence of well developed repulsive interactions in the neighbourhood of SC.

\begin{figure}[ht] 
\centerline{\includegraphics[width=1.1\hsize]{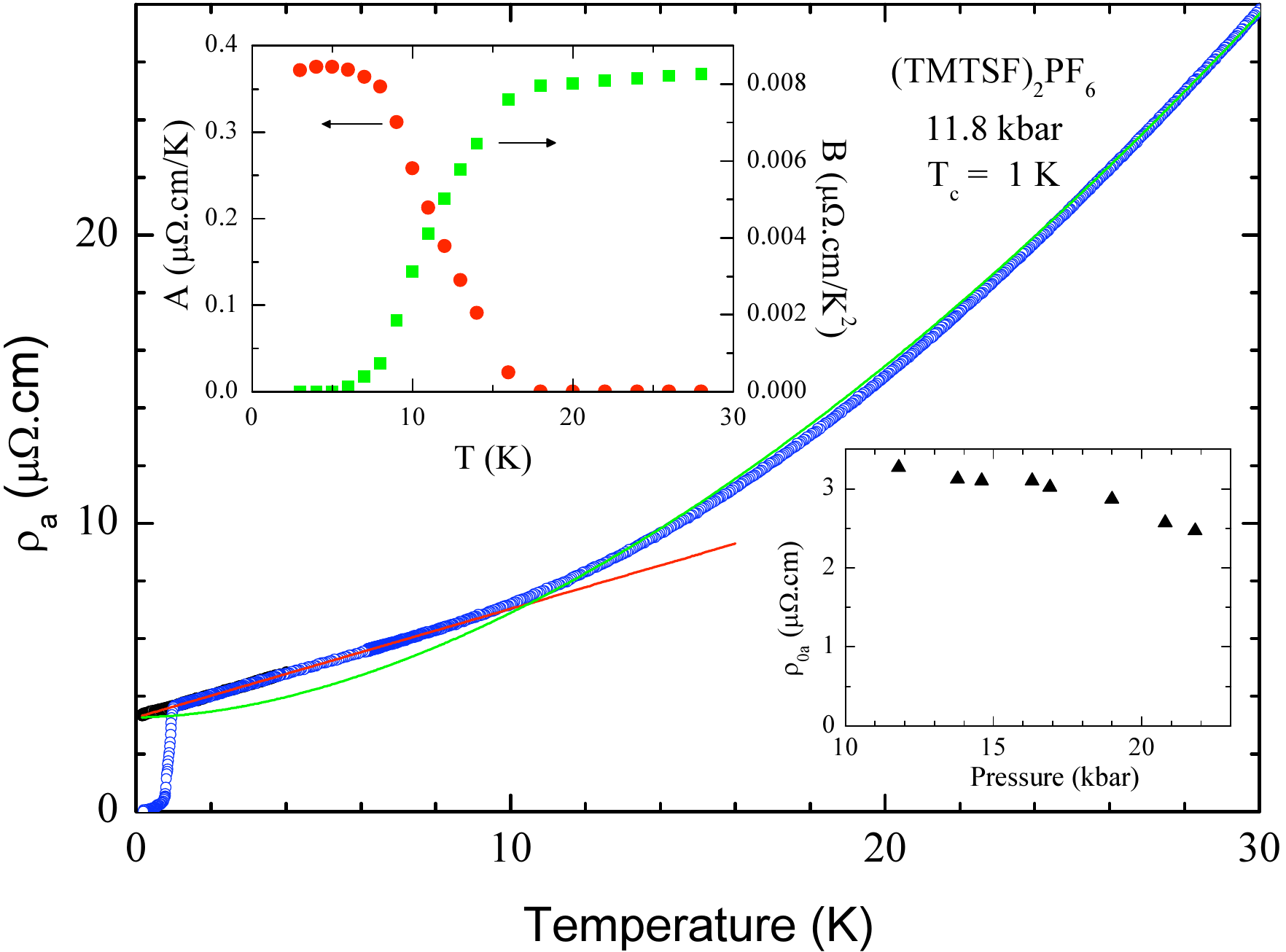}}
\caption{Temperature dependence of the longitudinal resistivity of \tmp6 at $P = 11.8$ kbar below 30K, at zero field and under $H=0.05$ T along $c^*$ in order to suppress SC.  Left insert: Temperature dependence of the $A$ and $B$ coefficients in the polynomial form $\rho(T) = \rho_0 + AT + BT^2ln(t_{\perp}/T)$ used to fit the resistivity data according to the sliding fit procedure described in the text. The linear and quadratic fits which are displayed in the main panel concern the low and high temperature extremes. Right insert: Pressure dependence of the residual resistivity $\rho_0$ deduced from the low temprature fit at different pressures.}
\label{Figure2bis.pdf}
\end{figure}

A reinvestigation of \tm2x up to pressures where SC is suppressed \textit{.i.e}. in the lower right corner of the phase diagram in figure\ref{Figure1.pdf} became  desirable  as it might clarify the long standing question about leading parameters governing the existence of organic superconductivity. This work was reported in ref\cite{Doiron09}, where it revealed the existence of a close correlation between a linear-in-temperature non Fermi liquid contribution  to the electrical resistivity,  at low temperature, $A$, and the SC \tc showing that both $A$ and \tc vanish at the same pressure.


\begin{figure}[ht]
\centerline{\includegraphics[width=0.9\hsize]{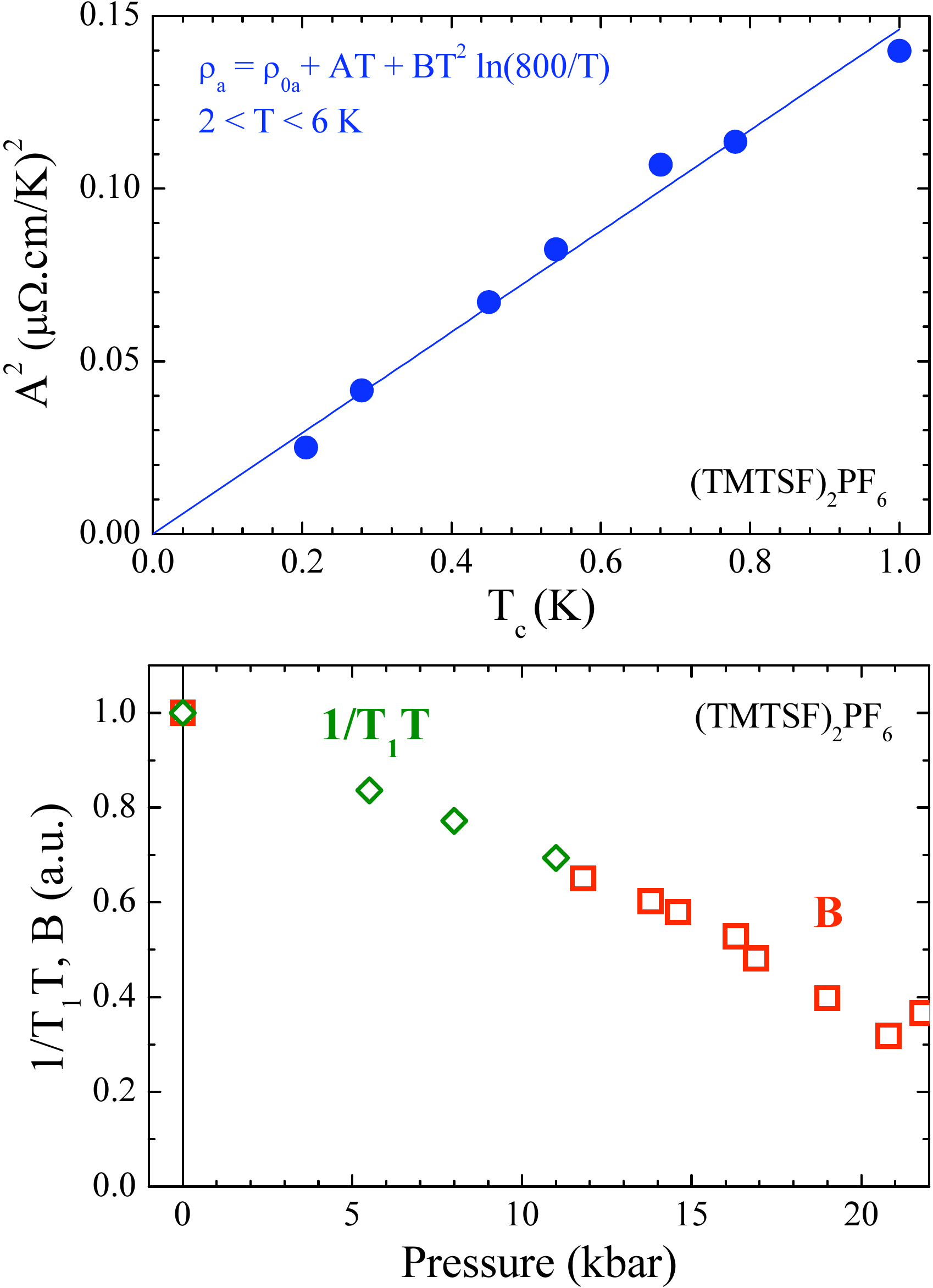}}
\caption{ (Top) Square of the $A$ coefficient, corresponding to the  value obtained at low temperature ($2<T<6$K) for the polynomial fit and the sliding fit procedure described in the text, against \tc onset. The line is a linear fit through the origin with the seven pressure points. (Bottom) Pressure dependence of the $B$ coefficient, corresponding to the value obtained at high temperature (26$<T<30$K) for the same fit, and of the  spin susceptibility measured under pressure \textit{via} NMR relaxation experiments\cite{Bourbonnais88}. Transport data at ambient pressure have been taken from ref\cite{Tomic91}. The relation $B\propto \chi^{2}$ is indicative of the Kadowaki-Woods relation  encountered in strongly correlated metals\cite{Kadowaki86}.}
\label{Figure3bis.pdf}
\end{figure}
The data analysis in\cite{Doiron09} focused on the low temperature regime (below 8 K). Here we examine the same data over a larger temperature range, up to 30~K.
In figure \ref{Figure2bis.pdf} we first show, in the right insert, the value of the residual resistivity $\rho_0$ determined by a linear extrapolation of the $\rho(T)$ data below 1~K, after suppression of SC by a small magnetic field (see\cite{Doiron09}). The smooth and weak   pressure dependence of $\rho_0$ is a good indication for  the quality of the data  of the  seven  pressure  runs. The pressure coefficient ($-2.5\%/kbar)$ can be attributed to the regular decrease   of the  effective mass under pressure\cite{Welber78,Jerome82}.

The resistivity of \tmp6   at a pressure of 11.8 kbar, close to the critical pressure where  $T_{SDW}\rightarrow  0$, is displayed in figure\ref{Figure2bis.pdf}. Below 8 K, it is strictly linear in temperature, with $\rho(T) = \rho_0 + AT$ .
Above 8 K, $\rho(T)$ acquires curvature, which can be captured by a  quadratic law $\rho(T) = \rho_0 +  BT^2$, keeping the same value for $\rho_0$. A small logarithmic factor actually improves the quality of the quadratic fit \cite{Gorkov98}. Note that this data just above the SDW critical pressure is very similar to the resistivity of the hole-doped cuprate Nd-LSCO at its quantum critical point\cite{Daou09}, where the temperature scale is approximately 10 times higher.


The resistivity can be analysed over a temperature domain from 1 to 30K, by performing  a two parameter fit of the data in figure\ref{Figure2bis.pdf} to the polynomial form $\rho(T) = \rho_0 + AT + BT^2ln(t_{\perp}/T)$  with a sliding temperature window of 4K and $t_\perp = 800$ K. The temperature dependence of the coefficients  $A$ and $B$ thus derived  is displayed in the left insert of figure\ref{Figure2bis.pdf}.


 We have applied this fitting procedure to our resistivity curves at all pressures and in figure\ref{Figure3bis.pdf}(top) we plot A, obtained in the 2 - 6 K window where $B$ is negligible, as a function of \tc, which is determined by the onset of SC in zero field. It is interesting to see that $A^2$ and \tc correlate fairly well  linearly and quite remarkably, that both vanish at the origin. In addition, the prefactor $B$ of the $T^2$ law  determined in the  $26-30K$ window,  and  the square of the electronic susceptibility ($\chi^2(q=0) \propto 1/T_{1}T$) follow the same pressure dependence, as shown in the bottom of figure\ref{Figure3bis.pdf}, which is reminiscent  of  the Kadowaki-Woods law observed in various strongly correlated metals\cite{Kadowaki86}.

In summary, a careful measurement of transport in \tmp6  up to 21.8 kbar has shown that a reasonable description of the longitudinal resistance up to 30 K or so can be obtained with a second order polynomial fit, provided $A$ and $B$ are allowed to evolve in temperature in a complementary way.
This study reveals clearly two extreme regimes. First, the high temperature regime where the $T^2$  scattering is dominant,  with  a logarithmic correction due to the 2D character  of the electron gas which can be taken as the signature of strong electron-electron Umklapp scattering\cite{Ziman60} in a 2D conductor\cite{Gorkov98}. The amplitude of this scattering is related to the spin susceptibility by a Kadowaki-Woods relation. Second, a low temperature regime where  linearity prevails in the temperature dependence of the resistivity, departing from the conventional Fermi liquid behaviour (even at one dimension) and  hinting at a tight correlation between such a scattering and the stability of SC. This behaviour can be regarded as a balance between two electron scattering processes adding according to the Mathiessen rule\cite{Ziman60}; one channel strictly linear up to 8K and vanishing above 15K or so and   one growing from zero around 8K and leveling off above 20K. 
\begin{figure}[ht]
\centerline{\includegraphics[width=1.1\hsize]{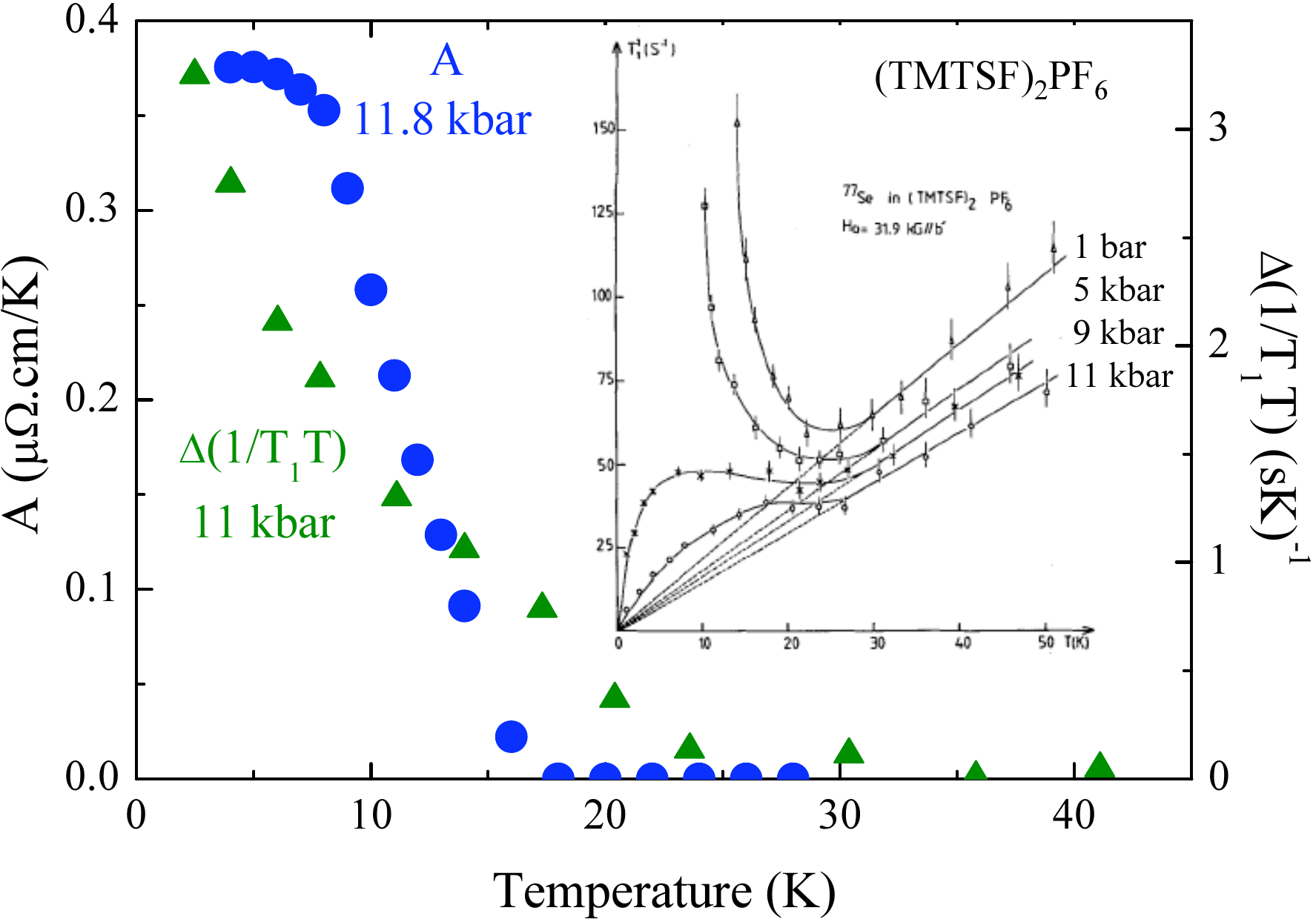}}
\caption{Temperature dependence of the fluctuation-induced relaxation $\Delta (1/T_{1}T)$ at 11 kbar from ref\cite{Bourbonnais88} and of the coefficient of the linear resistivity  ($A$) at 11.8 kbar. The insert shows the $^{77}$Se relaxation data at four different pressures where the bump of extra relaxation coming form AF fluctuations is clearly seen\cite{Bourbonnais88}. Similar results have been obtained in ref\cite{Brown08}.}
\label{Figure4.pdf}
\end{figure}

Note that the precise power with which $A$ scales with $T_c$ depends on the form of the fit and the fitting interval. In the fit performed here, where $\rho(T) = \rho_0 + AT + 
BT^2ln(800/T)$ over an interval from 2 to 6 K, we find $A \propto T_c^{0.5}$ (figure\ref{Figure3bis.pdf}). In ref\cite{Doiron09}, a fit of the form $\rho(T) = \rho_0 + AT + BT^2$ over an interval from $T_c$ to 8 K yields $A \propto T_c^{0.7}$. Performing the same fit over an interval from 0.1 to 4 K yields $A \propto T_c^{1.0}$\cite{Doiron09}. 
But in all cases, crucially, the onset of the linear term in the resistivity matches with the onset of SC with decreasing pressure, in other words, a pure $T^2$ Fermi liquid resistivity is only seen when Tc has been completely suppressed with increasing pressure. 

The relation between the non-Fermi liquid-like  scattering $A$ and  the behaviour of the $^{77}\mathrm{Se} $ spin-lattice relaxation rate measured previously in \tmp6 under pressure\cite{Bourbonnais88,Shinagawa07} is also enlightning.

When the contribution to the relaxation provided by AF fluctuations is extracted from the total relaxation rate one can compare its temperature dependence with that of the "linear" scattering. Such a comparison is shown in figure\ref{Figure4.pdf} where  the additional contribution to the ($q=0$) Korringa relaxation, $\Delta(1/T_{1}T)$, and $A$ are both seen to decrease with temperature.
Since transport and NMR experiments have not been conducted at exactly the same pressure a second best approach consists in  using the  transport data of the present measurements at 11.8 kbar and the 11 kbar NMR data available in the literature\cite{Bourbonnais88} (insert, figure\ref{Figure4.pdf}). Consequently, the comparison between  temperature dependencies of $A$ and $\Delta (1/T_{1}T)$ should not be taken at face value. This is only indicative of a common origin.

In conclusion, there exists in the low-temperature metallic phase of the salt \tmp6 close to the SDW order a correlation between SC and two properties of the metallic phase: (i) the non-Fermi liquid linear-$T$  resistivity as $T \to 0$ and (ii) the  contribution to the spin-lattice relaxation coming from the presence of antiferromagnetic fluctuations close to the SDW instability.  
Away from the SDW phase at elevated temperatures, the  resistivity acquires a $T^2$ curvature whose pressure dependence appear to correlate well with the Korringa law observed in NMR experiments when $T_c \to 0$. The correlation between non-Fermi-liquid resistivity and superconducting $T_c$ further suggests that anomalous scattering and pairing have a common origin\cite{Doiron09,Bourbonnais09,Bourbonnais09a}.

A similar correlation has been observed in \tmc under pressure\cite{Doiron09b} where NMR studies have also been performed. Furthermore, the  study of \tmc   has shown that organic superconductivity is governed by two control parameters: (i)  the intrinsic control parameter monitored by the intrinsic strength of interactions (pressure) and (ii) the pair breaking mechanism governed by the elastic electron life time which is quite influential in these spin-singlet superconductors\cite{Shinagawa07} with line nodes in the gap\cite{Joo04}.

We note that the phase diagram  of the iron-pnictide superconductors Ba(Fe$_{1-x}$Co$_x$)$_2$As$_2$,  with its adjacent semi-metallic SDW and superconducting phases\cite{Fang09,Chu09,Colombier09},  closely resembles that of \tm2x. The present data analysis supports the suggestion made in ref\cite{Doiron09} regarding a similar interpretation holding for   both  families of  materials.

We acknowledge C.Bourbonnais for his continous and helpful  interest  in this work and for the development of a theoretical framework presented at  the same conference\cite{Bourbonnais09a}. This work was supported by NSERC (Canada), FQRNT (Qu\'ebec), CFI (Canada), a Canada Research Chair (L.T.),  the Canadian Institute for Advanced Research and CNRS in France.



\end{document}